\documentclass{aa}  

\usepackage{graphicx}
\usepackage{txfonts}
\usepackage[colorlinks=true, citecolor=blue, linkcolor=blue]{hyperref}
\usepackage{color}
\usepackage[normalem]{ulem}
\usepackage{natbib}
\bibpunct{(}{)}{;}{a}{}{,} 

\begin{document}

   \title{Observational study of chromospheric heating by acoustic waves}

   \subtitle{}

   \author{V. Abbasvand
          \inst{1,2}\fnmsep\thanks{vahid.abbasvand.azar@asu.cas.cz}
          \and
          M. Sobotka\inst{1}
           \and
          M. \v{S}vanda\inst{1,2}
           \and
          P. Heinzel\inst{1}
           \and
          M. Garc\'{\i}a-Rivas\inst{1,2}
           \and
          C. Denker\inst{3}
           \and
          H. Balthasar\inst{3}
           \and
          M.~Verma\inst{3}
           \and
          I. Kontogiannis\inst{3}
                     \and
          J. Koza\inst{4}
                     \and
          D. Korda\inst{2}
                     \and
          C. Kuckein\inst{3}
      }

   \institute{Astronomical Institute of the Czech Academy of Sciences (v.v.i.),
Fri\v{c}ova 298, 25165 Ond\v{r}ejov, Czech Republic
\and
 Astronomical Institute of Charles University, Faculty of Mathematics and Physics,
V Hole\v{s}ovi\v{c}k\'{a}ch 2, 180 00 Praha 8, \\  Czech Republic
 \and
 Leibniz-Institut f\"{u}r Astrophysik Potsdam,
An der Sternwarte 16, 14482 Potsdam, Germany
\and
Astronomical Institute, Slovak Academy of Sciences,
059 60 Tatransk\'{a} Lomnica, Slovakia
             }

   \date{Received  June 2, 2020; accepted August 5, 2020}


\abstract
   {}
   {To investigate the role of acoustic and magneto-acoustic waves in heating the solar chromosphere, observations in strong chromospheric lines are analyzed by comparing the deposited acoustic-energy flux with the total integrated radiative losses.}
   {Quiet-Sun and weak-plage regions were observed in the Ca\,\textsc{ii} 854.2 nm and H$\alpha$ lines with the Fast Imaging Solar Spectrograph (FISS) at the 1.6-m Goode Solar Telescope (GST) on 2019 October 3 and in the H$\alpha$ and H$\beta$ lines with the echelle spectrograph attached to the Vacuum Tower Telescope (VTT) on 2018 December 11 and 2019 June 6. The deposited acoustic energy flux at frequencies up to 20~mHz was derived from Doppler velocities observed in line centers and wings. Radiative losses were computed by means of a set of scaled non-LTE 1D hydrostatic semi-empirical models obtained by fitting synthetic to observed line profiles.}
   {In the middle chromosphere \mbox{($h$ = 1000--1400 km)}, the radiative losses can be fully balanced by the deposited acoustic energy flux in a quiet-Sun region. In the upper chromosphere ($h > 1400$~km), the deposited acoustic flux is small compared to the radiative losses in quiet as well as in plage regions. The crucial parameter determining the amount of deposited acoustic flux is the gas density at a given height.}
   {The acoustic energy flux is efficiently deposited in the middle chromosphere, where the density of gas is sufficiently high. About 90\% of the available acoustic energy flux in the quiet-Sun region is deposited in these layers, and thus it is a major contributor to the radiative losses of the middle chromosphere. In the upper chromosphere, the deposited acoustic flux is too low, so that other heating mechanisms have to act to balance the radiative cooling.}


   \maketitle
%

\section{Introduction}

The solar chromosphere is hotter than the photosphere and the increase of temperature in semi-empirical models of the chromosphere cannot be explained by radiative heating. The strong spectral lines of neutral hydrogen, Ca\,\textsc{ii}, and Mg\,\textsc{ii} are the most important lines for studying the released radiative energy from the chromosphere \citep{carlsson2019}. The cores of these spectral lines are formed under non-LTE conditions, where departures from the local thermodynamical equilibrium (LTE) are important. In the quiet Sun, \cite{Vernazza1981} integrated the radiative losses over the height of the chromosphere, obtaining 4600~W\,m$^{-2}$. In active regions, these losses are higher by a factor of 2--4 \citep{Withbroe1977}. They have to be balanced by an energy input supplied by various heating mechanisms.

There are several candidates of chromospheric heating mechanisms, which can be classified in two main competitive groups \citep[see][ for a review]{Jess2015}: (1) heat release related to two groups of magnetic field lines with opposite directions and their reconnection, which creates local electric current sheets to release the magnetic energy \citep{Rabin1984, Testa2014}; (2) dissipation of the energy of upward-propagating magneto-acoustic waves. These waves stem from the acoustic waves generated in the upper convection zone because of turbulent motions \citep[e.g.,][and references therein]{Aschwanden2001,Zaqarashvili2009,Kayshap2018}. In this paper, we focus to the latter mechanism.

The magnetic field affects the propagation of acoustic waves. The acoustic waves undergo the mode conversion of magneto-acoustic waves on equipartition surfaces in the upper photosphere, where the Alfv\'en velocity is equal to the sound speed. The inclined magnetic field then facilitates also the propagation of the waves with frequencies below the acoustic cut-off frequency \mbox{$\nu_{\rm ac}$ = 5.2 mHz} \citep{Bel1977} into the upper atmosphere through so-called magnetic portals  \citep{Jefferies2006,Stangalini2011,Konto2014,Konto2016}. 

The physical properties of the solar chromosphere including its dynamics, energetics, and the formation of spectral-line features have been explored with significant efforts, using both theoretical simulations and observational analysis. Theoretical evidence shows that acoustic waves can heat the intranetwork (i.e., non-magnetic) regions on the Sun \citep{Ulmsch2003}. Based on simulations, \cite{Carlsson1992} reported that when the acoustic wave dissipation is fully employed, the Ca\,\textsc{ii}\,H line behavior is very similar to the observed one. \cite{Cuntz2007} discussed the physical nature of acoustic heating in the solar chromosphere using time-dependent 1D simulations to calculate wave energy fluxes. They claimed that high-frequency acoustic waves are sufficient to heat the intranetwork regions in the solar chromosphere.

According to the work of \cite{Bello2010a, Bello2010b}, the total acoustic flux in quiet regions calculated from observed Doppler velocities at the height of 500--600 km is approximately 2000~W\,m$^{-2}$ and in the photosphere ($h = 250$~km) it reaches 6500~W\,m$^{-2}$. Its spatial distribution in the field of view is strongly inhomogeneous. \cite{Beck2009} estimated acoustic energy fluxes from velocity oscillations in the chromospheric Ca\,\textsc{ii} H line core and in several photospheric lines blended in the wings of that line. They found that the acoustic flux of 1000~W\,m$^{-2}$ at the heights 800--1200~km was insufficient to maintain the temperature stratification of chromospheric semi-empirical models above the temperature minimum.

The dissipation of (magneto)acoustic waves is generally time-dependent. \cite{Carlsson1995, Carlsson1997} applied time-dependent 1D radiation-hydrodynamic models to compute the propagation of acoustic waves in the non-LTE regime. They found that acoustic shocks cause short time intervals of high temperature, however, the average temperature of the chromosphere continuously decreases with height. This result disagrees with the fact that emission in chromospheric lines exists everywhere and all the time \citep{Carlsson1997, Curdt1998, Kalkofen1999}. The discrepancy might be solved by a multi-dimensional approach, but the difficulties connected with multi-dimensional time-dependent non-LTE modeling, namely extremely high computational requirements, do not allow us to represent the observed chromospheric structures directly this way \citep{Carlsson2012}. A review on radiation-hydrodynamic models of the solar atmosphere was recently published by \citet{Leenaarts2020}.

\defcitealias{sobotka2016}{Paper~I}
\defcitealias{Abbasvand2020}{Paper~II}

An alternative is the stationary approach, which uses time-averaged atmospheric parameters of 1D semi-empirical hydrostatic models. The energy flux of (magneto)acoustic waves is calculated from power spectra of Doppler velocities obtained during a time interval of several tens of minutes. It is a time-averaged quantity, which cannot be compared with instantaneously radiated energy but with time-averaged radiative losses derived from the semi-empirical models. This approach was used by \citet[][hereafter Paper~I]{sobotka2016}  who, analyzing the velocities measured from the \mbox{Ca\,\textsc{ii} 854.2 nm} line, showed that the acoustic energy flux deposited in the middle chromosphere provides a remarkable source of energy required to balance the local radiative losses.

\citet[][hereafter Paper~II]{Abbasvand2020} recently introduced a new grid of semi-empirical 1D hydrostatic models, which were obtained by scaling the temperature and column mass of six initial models VAL A--F \citep{Vernazza1981}. This grid describes most of the solar features at moderate spatial resolution ($1''$), including both the quiet and active Sun, and enables us to scrutinize the hypothesis of chromospheric heating by acoustic and magneto-acoustic waves. Using the data analyzed already in Paper I, they also demonstrated that in the middle chromosphere, the contribution of the deposited (magneto)acoustic flux energy to the released radiative energy in the quiet Sun is approximately 30--50\,\%. In a weak plage, it is 50--90\,\%, and values above 70\,\% correspond to locations where the magnetic field is inclined more than 50$^{\circ}$ with respect to the solar surface normal. The cadence of observations was 52 s, limiting the maximum detectable wave frequency to 9.6 mHz.

In Papers I and II, a single region on the Sun was examined using only one spectral line formed in the middle chromosphere. More regions have to be studied in several chromospheric lines to expand the range of heights in the chromosphere, arrive at plausible results, and to validate the working hypothesis. Therefore, in the present work, we set out to estimate the contribution of (magneto)acoustic waves to chromospheric heating by means of observations of three different, quiet and active, regions in combinations of three chromospheric lines H$\alpha$, H$\beta$, and \mbox{Ca\,\textsc{ii} 854.2 nm} with a raster scan time of about 25~s. This cadence provides a maximum detectable wave frequency of approximately 20~mHz, making it possible to include also high-frequency waves.


\section{Observations and data processing} \label{sec:observation}

\begin{table*}\centering
\caption{Parameters of the VTT and BBSO data sets.}  \label{tab:VTT_table}
\begin{tabular}{llllcccc}
\hline\hline
Instrument & \multicolumn{2}{c}{VTT/echelle}   & GST/FISS\\
Data set  & 2018 December 11 &  2019 June 6 & 2019 October 3\\
\hline
Time & 11:42 UT &  11:00 UT & 17:46 UT\\
Target & Weak AR\tablefootmark{*} - pore & Weak AR\tablefootmark{*} - plage & Quiet Sun \\
Scanned lines & H$\alpha$ 656.28 nm & H$\alpha$ 656.28 nm & H$\alpha$ 656.28 nm \\
& H$\beta$ 486.13 nm  & H$\beta$ 486.13 nm & Ca\,\textsc{ii} 854.2 nm \\
Coordinates & $-423''$E, 147$''$N  & $-416''$E, 122$''$N  &  0$''$E, 0$''$N\\
Cosine of heliocentric angle & $\mu = 0.88$ & $\mu = 0.89$ & $\mu = 1.00$ \\
No. of spectral points &  2004 (H$\alpha$)   &  334 (H$\alpha$) &  512 (H$\alpha$)  \\
 & 1560 (H$\beta$)  & 334 (H$\beta$) & 502 (Ca\,\textsc{ii} 854.2 nm)\\
Wavelength spacing & 0.41 pm (H${\alpha}$) & 1.66 pm (H$\alpha$) & 1.9 pm (H$\alpha$) \\
 & 0.3 pm (H$\beta$)  &  1.21 pm (H$\beta$) & 2.6 pm (Ca\,\textsc{ii} 854.2 nm) \\
Wavelength range & 822 pm (H$\alpha$)  & 555.3 pm (H$\alpha$) & 972.8 pm (H$\alpha$) \\
& 468 pm (H$\beta$)  & 403.3 pm (H$\beta$)  & 1305.2 pm (Ca\,\textsc{ii} 854.2 nm)  \\
Field of view & 14\farcs4 $\times$ 118\farcs8  & 50\farcs4 $\times$ 180$''$ & $20''\times 41''$\\
Region of interest & 14\farcs4 $\times$ 45\farcs54  & 50\farcs4 $\times$ 126$''$ & 13\farcs12 $\times$ 38\farcs4 \\
Pixel size &  0\farcs36$\times$0\farcs18 & 0\farcs36$\times$0\farcs36 & 0\farcs16$\times$0\farcs16 \\
Exposure time & 300 ms   & 80 ms & 60 ms\\
Raster scan time &  25 s  & 25 s & 22.9 s\\
No. of spectral scans &  60  & 200 & 210\\
\hline
\end{tabular}
\tablefoot{\tablefoottext{*}{Active region}}
\end{table*}

\subsection{Data sets} \label{sec:Dataset}

The observations were performed during three campaigns at two different instruments, both being long-slit scanning spectrographs:

\begin{itemize}

\item The echelle spectrograph attached to the 70-cm German Vacuum Tower Telescope \citep[VTT,][]{Schroeter1985,vonderLuhe1998}, operating at the Observatorio del Teide, Tenerife, Spain -- data sets on 2018 December~11 and 2019 June~6.

\item The Fast Imaging Solar Spectrograph \citep[FISS, ][]{Chae2013} attached to the 1.6-m Goode Solar Telescope \citep[GST,][]{Caoetal2010} at the Big Bear Solar Observatory (BBSO), California, USA -- data set on 2019 October 3.
\end{itemize}

\noindent
The most important data set was obtained with FISS at GST in a service mode, because simultaneous observations of the spectral lines Ca\,\textsc{ii} 854.2 nm and H$\alpha$ make it possible to measure the deposited acoustic flux in the middle and in the upper chromosphere. The target was a quiet-Sun area located exactly at the center of the solar disc. Some small magnetic elements were dispersed in this area. The advantage of FISS is its high spectral and spatial resolution combined with short exposure times that facilitate fast scanning of the observed area on the Sun. The seeing was good and stable during the 80-minute period of observation that was selected for further analysis. With the aid of the adaptive-optics system installed at GST, high-order correction of atmospheric seeing was provided within an isoplanatic patch, with a gradual roll off of the correction at larger distances \citep{Shumko2014}. Additional observations were taken with the Near Infra-Red Imaging Spectropolarimeter \citep[NIRIS,][]{Cao2012} to obtain information about the magnetic field. Due to some calibration issues, it was not possible to invert the spectropolarimetric data.

Two other data sets, obtained using the echelle spectrograph at VTT, allowed us to study only the upper chromosphere. The first one was acquired on 2018 December 11 and the second on 2019 June 6. In both cases, weak magnetic regions near disc center were scanned where weak chromospheric plages and small pores were present. The spectra were recorded simultaneously in the H$\beta$ and H$\alpha$ lines of neutral hydrogen. The seeing conditions were average during the data-acquisition period of 25 minutes on December 11 and good and stable during the 83-minute period on June 6. The Kiepenheuer Adaptive Optics System \citep[KAOS,][]{vdLuehe2003} working at a wavelength of 500\,nm indicated a Fried parameter of \mbox{5--6 cm} on December 11 and \mbox{7--9 cm} on June~6.

\begin{figure}[t]\centering
\includegraphics[width=3.3in]{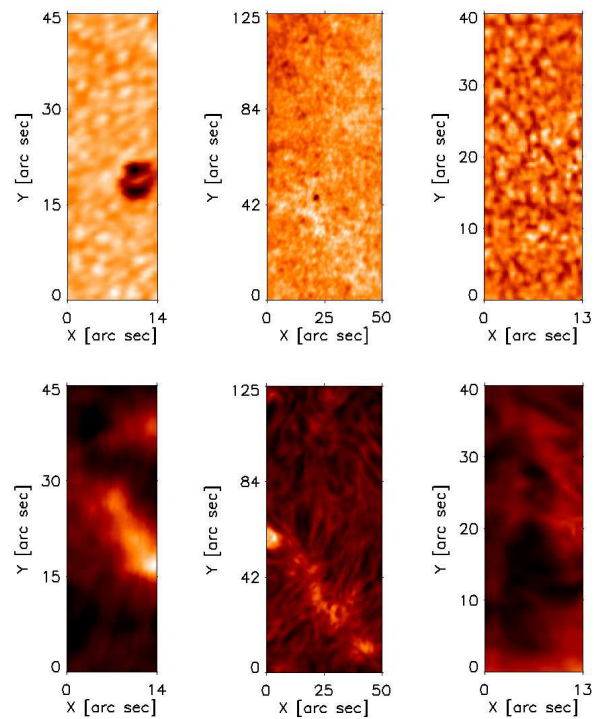}
\caption{Maps of continuum intensity ({\it top}) and H$\alpha$ line-center intensity ({\it bottom}) corresponding to data sets of 2018 December 11 ({\it left}), 2019 June 6 ({\it middle}), and 2019 October 3 ({\it right}).
\label{fig:fig1}}
\end{figure}

The raw data were calibrated using standard dark- and flat-field procedures and the FISS data were compensated for image rotation in the coud\'e focal plane of the telescope. In the December 11 VTT data set, the spatial sampling along the spectrograph slit is finer than the scanning step, so that the spatial pixels are rectangular. The relevant characteristics of all data sets are summarized in Table \ref{tab:VTT_table}. Regions of interest (ROI) of all data sets in the continuum and H$\alpha$ line center are displayed in Figure~\ref{fig:fig1}.

Additional polarimetric observations in the photospheric line Fe\,\textsc{i} 1564.9~nm were acquired on December~11 with the GREGOR Infrared Spectrograph \citep[GRIS,][]{collados2012} attached to the \mbox{1.5-m} telescope GREGOR \citep{Schmidt2012} at the Observatorio del Teide. These polarimetric data were inverted with the Stokes inversion code based on response functions \citep{RuizCobo1992}. The spatial resolution of the resulting magnetic-field vector map was 0\farcs 5.

\subsection{Data processing} \label{sec:Dataprocessing}

The Doppler velocities were measured for two purposes: (1) to remove Doppler shifts from the spectroscopic observations to calculate time-averaged line profiles and (2) to obtain purely chromospheric Doppler velocities required for the calculation of acoustic fluxes. In both cases, we used the \texttt{bisec\_abs\_lin.pro} routine included in the Leibniz-Institut f\"ur Sonnenphysik (KIS) IDL library. This routine returns line shifts at pre-selected intensity levels in the line profile (a bisector calculation) and the shift of the line centre, derived from a parabolic fit around the minimum intensity. The shifts were converted into the Doppler velocities assuming that the quiet-Sun regions within the field of view are at rest on average.

To remove Doppler shifts, we averaged bisector positions at five intensity levels from 40\,\% to 70\,\%, where 0\,\% corresponds to the line-center intensity and 80\,\% to the line-wing intensity near the continuum. The whole line profile was then moved back by the obtained average shift. To obtain the chromospheric velocities, we used the line-center shifts
of H$\alpha$ and Ca\,\textsc{ii} 854.2 nm lines and bisector shifts of H$\beta$ and Ca\,\textsc{ii} at line-profile intensity levels corresponding respectively to wavelength distances of $\pm 18$~pm and $\pm 13$~pm from the line center, which were found from contribution functions (see Section~\ref{sec:Modelatmosphere}).

Mean profiles of all spectral lines were obtained by time-averaging over the observing period of each series of the observed profiles with removed Doppler shifts. They are used to find the most appropriate semi-empirical models at each location in ROIs (Section~\ref{sec:Modelatmosphere}).

The photospheric magnetic-field strength and inclination are necessary for the acoustic-flux calculation. Because this information was not available for two (June 6 and October 3) of the three data sets, we decided to retrieve it for all the data sets from an independent common source. The magnetic-field maps are a data product from the Helioseismic and Magnetic Imager \citep[HMI,][]{HMI_Schou2012} onboard the Solar Dynamics Observatory \citep[SDO,][]{SDO_Pesnell2012}. The HMI Vector Magnetic Field Pipeline \citep{HMI_Bpipeline_Hoeksema2014} obtains the magnetic-field vector from Stokes parameters. The Stokes parameters are measured at six wavelength positions across the Fe\,\textsc{i} 617.3\ nm line for the full disc and are inverted using the Very Fast Inversion of the Stokes Vector code \citep[VFISV,][]{VFISV_Borrero2011,VFISV_Centeno2014}. VFISV solves the radiative-transfer equation considering a Milne-Eddington atmosphere.
The azimuth is disambiguated according to the existence of a strong field. In regions with strong magnetic field and their surroundings, a minimum-energy method is applied \citep{Metcalf1994,Leka2009}. In regions with weak magnetic field, the azimuth that makes the magnetic-field vector more similar to the potential-field vector is selected. The inverted vector magnetic-field angular parameters, i.e., the azimuth and inclination, are referred to the line-of-sight (LOS) frame and they are converted to the local reference frame (LRF) using the AZAM code \citep{Lites1995}. The spatial resolution of HMI magnetic-field vector maps is 1$''$.

\begin{figure}[t]\centering
\includegraphics[width=3.4in]{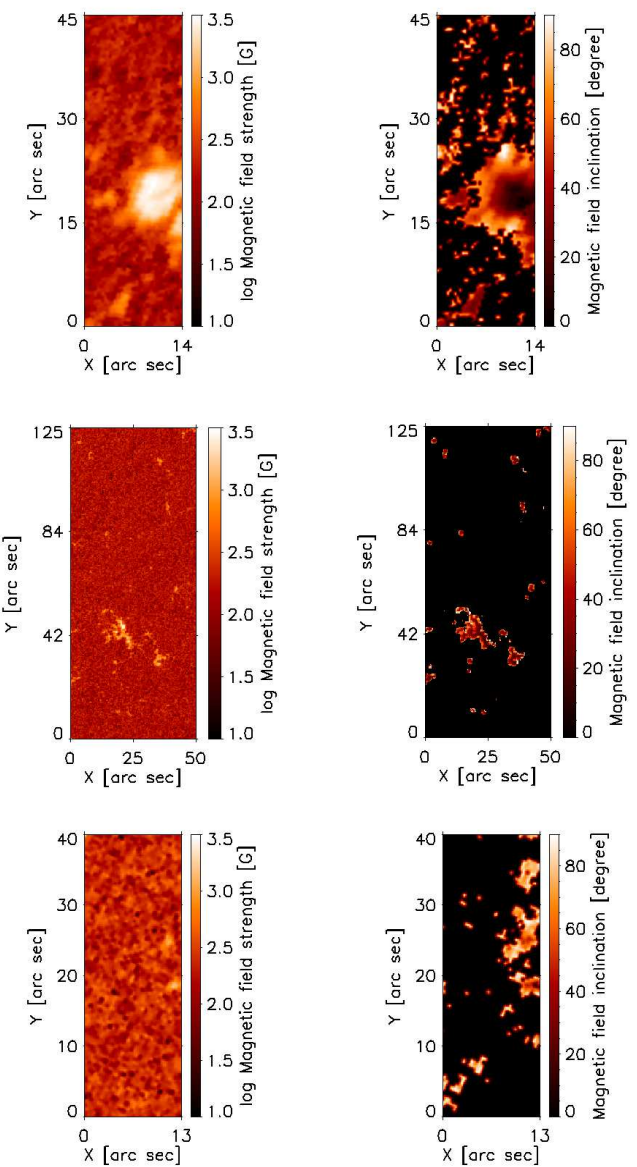}
\caption{Maps of photospheric magnetic-field strength and inclination in the ROIs for December 11 ({\it top}), June 6 ({\it middle}), and October 3 ({\it bottom}).
\label{fig:fig2}}
\end{figure}

The HMI data sets were obtained on December 11 at 12:00\ UT, June 6 at 11:24\ UT, and October 3 at 18:12\ UT and the full-disc maps of continuum intensity, magnetic-field strength, and inclination were cropped to match the ROIs. When the magnetic-field strength is low ($|B| < 125$~G in the case of HMI), the inversion code returns unreliable LOS inclinations $\theta \approx 90^{\circ}$ because of noisy Stokes $Q$ and $U$ signals. To remove these unreliable values, we applied a mask that set $\theta$ to zero in the regions where $|B|$ was below this limit. Figure \ref{fig:fig2} shows the resulting masked maps of photospheric LRF inclination and maps of $|B|$ in logarithmic scale for all data sets. The effect of masking was discussed in Paper II, where we showed that unreliable values of $\theta$ led to overestimated acoustic fluxes. 

To align the spectroscopic observations with magnetic-field maps, we need to compare pores and granulation observed in VTT and GST continua (reference images) to those in HMI continuum images (transformed images). These features can be identified in both types of continuum images. The transformed an reference images are inputs for the following semi-automatic spatial alignment procedure.

First, a pair of co-temporal images is inspected visually by the routine \texttt{setpts.pro} (or \texttt{setpts\_roi.pro}) implemented within the IDL SolarSoft System \citep[SSW,][]{FreelandHandy1998}. The routine serves for an interactive definition of common points in two images. Then the set of reference points is used to calculate a linear transformation, which maps one image onto another, by the SSW function \texttt{caltrans.pro}. Finally, the initial transform parameters are entered into the SSW function \texttt{auto\_align\_images.pro} by T.~Metcalf, which looks for the best spatial alignment of an image pair based on their cross-correlation. Five cross-alignment parameters, i.e., [$x,y$] shifts, [$x,y$] magnification factors, and rotation angle, which yield the maximum cross-correlation, are inferred in two subsequent calls of the function. In the first call, the downhill simplex method \citep{Pressetal1992}, implemented as the \texttt{amoeba.pro} routine, is applied. Its results are used as initial parameters for the second call, using the more robust Powell method \citep{Pressetal1992}. By this two-step procedure, a satisfactory spatial alignment between the co-temporal image pairs GST -- HMI and VTT -- HMI (June and December) are achieved with the cross-correlation coefficients of 0.90, 0.92 and 0.85, respectively.


\section{The method} \label{sec:method}

The applied method calculates the acoustic energy flux deposited in a chromospheric layer between two reference geometrical heights and compares it with radiative losses computed from model atmospheres and integrated over the same height range. The reference heights are determined with regard to typical formation heights of the observed spectral lines.

\subsection{Model atmospheres} \label{sec:Modelatmosphere}

The initial semi-empirical models VAL A--F include the column mass $m$, optical depth $\tau_{500}$ at 500 nm, temperature $T$, microturbulent velocity $\upsilon_{\rm t}$, hydrogen density $n_{\rm H}$, electron density $n_{\rm e}$, total pressure $P_{\rm {tot}}$, gas pressure $P_{\rm g}$, and density $\rho$ as a function of geometrical height $h$. To approximate the physical conditions in the ROIs, these initial models are scaled by changing two free parameters $p_{\rm T}$ and $p_{\rm m}$ that define the $T$ and $m$ stratifications \citepalias[see][for details]{Abbasvand2020}. These stratifications are used as inputs in the non-LTE radiative-transfer code based on the Multi-level Accelerated Lambda Iterations technique \citep[MALI,][]{Rybicki1991, Rybicki1992} with standard partial frequency redistribution (PRD). The hydrogen version of the code (with PRD in Lyman lines) computes the ionization structure and populations of hydrogen levels using a 5-level plus continuum atomic model. It re-computes the complete scaled atmospheric models and calculates synthetic profiles of the H$\alpha$ and H$\beta$ lines. As a next step, the calcium version of the MALI code (with PRD in the K and H lines) is used to get synthetic profiles of the Ca\,\textsc{ii} 854.2~nm line. Again, a 5-level plus continuum atomic model is used. For each of the initial VAL models, a grid of 2806 scaled models is computed using a combination of 61 and 46 possible values of $p_{\rm T}$ and $p_{\rm m}$, respectively. In total, a grid of 16836 models, parameterized by the initial model selection, $p_{\rm T}$, and $p_{\rm m}$, is available \citepalias{Abbasvand2020}.

The initial models are assigned to different areas in ROIs in accordance with the line-core intensity of the observed chromospheric lines. Then, a scaled model from the grid, which provides the best match of the synthetic to the local time-averaged observed profiles, is selected at each position within the ROIs. The best match means that the sum of squared differences between the synthetic and observed profiles of all lines under study (merit function) is at minimum.

\begin{figure}[t]\centering
\includegraphics[width=3.3in]{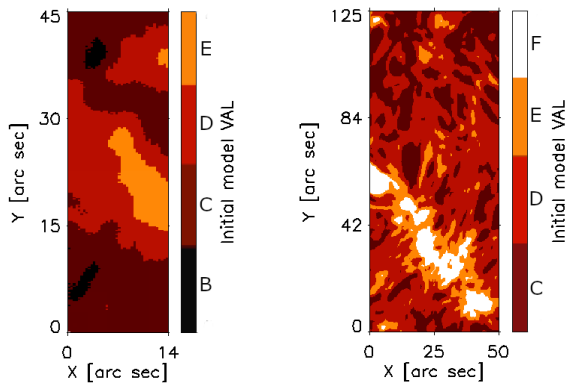}
\caption{Initial models for December 11 ({\it left}) and June 6 ({\it right}). The October 3 data set has only one initial model, VAL C.
\label{fig:fig3}}
\end{figure}

The optimal selections of initial models for the ROIs of the December, June, and October data sets are VAL B--E, VAL C--F, and VAL C, respectively. Figure~\ref{fig:fig3} shows the selection maps of the initial models in the ROIs of the December and June data sets. Finally, the December, June, and October ROIs are described by 482, 657, and 59 different scaled models, respectively.

\begin{figure}[t]\centering
\includegraphics[width=3.4in]{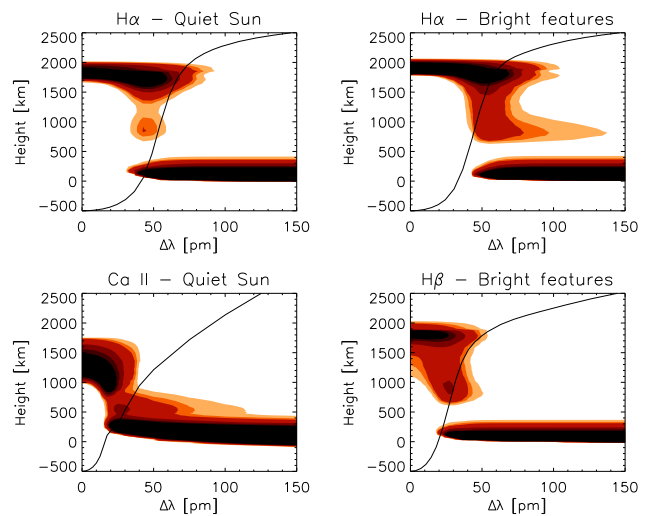}
\caption{Contribution functions of the H$\alpha$, H$\beta$, and Ca\,\textsc{ii} 854.2 nm lines for two typical models derived from our observations -- quiet Sun ({\it left}) and bright chromospheric features ({\it right}). The colors change from light orange to black with increasing values of the contribution function. Black lines show corresponding synthetic line profiles.
\label{fig:fig4}}
\end{figure}

Physical quantities, which are used to calculate acoustic fluxes, are retrieved from the models at reference heights that are common for all models in a ROI. These heights also delimit integration ranges of radiative losses. The reference heights are selected with regard to formation heights of the observed spectral lines, properly speaking, formation heights of parts of line profiles utilized to measure Doppler velocities.
The formation heights can be obtained from a contribution function $C_\nu(h)$, which describes the contribution of different atmospheric layers to the emergent intensity of radiation $I_{\nu}$ at the considered frequency $\nu$ \citep[e.g.,][]{Gurtovenko1974},
\begin{equation}
I_{\nu} = \int C_{\nu}(h) \, {\rm d}h.
\end{equation}
In quiet- and active-Sun atmospheric models, $C_\nu(h)$ of H$\alpha$, H$\beta$ and Ca\,\textsc{ii} 854.2 nm lines are computed by the non-LTE radiative-transfer codes described above.
Contribution functions computed from two typical models that characterize the quiet Sun (October data set) and bright chromospheric features (December and June data sets) are plotted for different wavelength distances $\Delta\lambda$ from the line center and different heights $h$ in Figure~\ref{fig:fig4}.
We can see that the central parts of line profiles are formed largely in the chromosphere, each of them in different ranges of heights.

The typical formation height can be calculated as a mean $h$ weighted by $C_\nu(h)$ at a given $\Delta\lambda$. We obtained typical formation heights using all models in each ROI and found the most frequent ones. The H$\alpha$ line center is formed typically at 1800~km in the quiet-Sun atmosphere and at 1900~km in bright chromospheric features. For the Ca\,\textsc{ii} 854.2 nm line in the quiet Sun, the typical formation height of the line center is 1400~km and the inner wings at $\Delta\lambda$\,=\,$\pm$13~pm are formed at 1000~km. From contribution functions computed at this $\Delta\lambda$ for each pixel in the October ROI we find that only 4\,\% of pixels have a contribution from the photosphere, which, however, is smaller than 1\,\%. The H$\beta$ line is more problematic. Its inner wings at $\Delta\lambda$\,=\,$\pm$18~pm are formed typically at 1600~km in bright chromospheric features (Figure~\ref{fig:fig4}) but 18\,\% of pixels in the December and June ROIs have photospheric contributions larger than 20\,\% at this wavelength.
The selection of $\Delta\lambda$ for H$\beta$ and Ca\,\textsc{ii} was intended to reach the deepest possible chromospheric layers with a minimum contribution from the photosphere. The most frequent typical formation heights, stated above, were accepted as the reference heights for the calculation of radiative losses and deposited acoustic fluxes.

The scatter of typical formation heights around the reference height is characterized by standard deviations of 20~km in the middle and upper chromosphere of the quiet-Sun region (October 3 data set, Ca\,\textsc{ii} and H$\alpha$) and 100~km in the upper chromosphere of weak active regions with a wide variety of models (December 11 and June 6 data sets, H$\beta$ and H$\alpha$). Although the latter value is large, it does not introduce any error into physical quantities and radiative losses computed from the models, because they refer to fixed reference heights. The difficulty arises in Doppler velocities, which are sometimes measured at heights different from the reference ones. We analyzed root-mean-squared (rms) Doppler velocities $\sigma_{\upsilon}$ measured at different $\Delta\lambda$, corresponding to the formation-height uncertainty of $\pm 100$~km, in the H$\alpha$ and H$\beta$ profiles of the December and June data sets. The relative difference $\Delta \sigma_{\upsilon} / \sigma_{\upsilon}$ was approximately of 7\,\% in the central parts of profiles formed purely in the chromosphere and increased by a factor of two at $\Delta\lambda$ with photospheric contribution. Because the acoustic energy flux is proportional to $\sigma_{\upsilon}^2$, its relative error was estimated to $\Delta \sigma_{\upsilon}^2 / \sigma_{\upsilon}^2 \simeq 2 \Delta \sigma_{\upsilon} / \sigma_{\upsilon} \simeq 15$\,\% and 30\,\%, respectively. For the October 3 data set, the relative error of the acoustic flux is 3\,\%.

\begin{table*}\centering
\caption{Reference heights $h_{\rm ref}$, rms Doppler velocities $\sigma_{\upsilon}$, mean densities  $\langle \rho \rangle$, and mean gas pressures $\langle P_{\rm g} \rangle$ at the reference heights.}
\label{tab:table2}
\begin{tabular}{lccll}
\hline\hline
Data set & $h_{\rm ref}$~[km] &  $\sigma_{\upsilon}$~[km s$^{-1}$]  & $\langle \rho \rangle$~[kg m$^{-3}$] & $\langle P_{\rm g} \rangle$~[Pa]\\
\hline
Dec 11  & 1600  & 0.83  & $(2.96 \pm 1.57) \times 10^{-9}$  & $0.158 \pm 0.086$\\
        & 1900  & 1.11  & $(6.27 \pm 4.01) \times 10^{-10}$ & $0.070 \pm 0.028$\\
\hline
Jun 6   & 1600  & 0.79  & $(3.18 \pm 1.32) \times 10^{-9}$  & $0.173 \pm 0.070$\\
        & 1900  & 1.64  & $(6.13 \pm 3.87) \times 10^{-10}$ & $0.087 \pm 0.023$\\
\hline
Oct 3   & 1000  & 1.62  & $(5.64 \pm 0.55) \times 10^{-8}$  & $2.202 \pm 0.241$\\
        & 1400  & 1.65  & $(5.37 \pm 0.71) \times 10^{-9}$  & $0.225 \pm 0.033$\\
        & 1800  & 2.31  & $(6.43 \pm 0.29) \times 10^{-10}$ & $0.065 \pm 0.012$\\
\hline
\end{tabular}
\end{table*}

The net radiative cooling rates (radiative losses) characterize the energy released by radiation from the chromosphere. These also indicate the amount of non-radiative heating that maintains the semi-empirical temperature at a given height in the solar atmosphere. To calculate them, we use the scaled models assigned to each pixel in a ROI. For each model, they are computed for the main contributors in the solar chromosphere, namely, the lines Ca\,\textsc{ii} H \& K, the Ca\,\textsc{ii} infrared triplet, Mg\,\textsc{ii} h \& k, hydrogen Lyman and Balmer lines, and hydrogen continua, using our non-LTE radiative transfer codes MALI. The total net radiative cooling rates, a sum of the main contributors, are integrated over the geometrical height in the range between the reference heights for each pixel in the ROIs.

\subsection{Deposited acoustic flux} \label{sec:Deposited}

Oscillations at each position in the ROIs are studied using standard Fourier analysis of the time series of Doppler maps measured in the line centers of H$\alpha$ and Ca\,\textsc{ii} 854.2 nm and at $\Delta\lambda$ = $\pm$18 pm and $\Delta\lambda$ = $\pm$13 pm from the line centers of H$\beta$ and Ca\,\textsc{ii} 854.2 nm, respectively. The cadence of the time series is around 25~s and thus the maximum detectable frequency 20~mHz is sufficient to investigate the power spectra of velocity oscillations in the quiet-Sun and magnetic regions in the chromosphere. The method of power-spectra calculation and calibration in absolute units is described by \citet{Rieutord2010}.

The acoustic-energy flux at a given height in the chromosphere is estimated following the method of \cite{bello2009}, assuming that (i) the acoustic waves propagate upwards and (ii) the inclination of magnetic field along the track of acoustic waves is equal to that measured in the photosphere.
It can be formulated as
\begin{equation}
\begin{split}
F_{\rm ac}(\nu) = \rho \, P_\upsilon(\nu) \, \upsilon_{\rm gr}(\nu) / T\!F(\nu),  
\label{equ:2}
\end{split}
\end{equation}
where $\rho$ is the gas density at the given height, $P_\upsilon$ the spectral-power density derived from the Doppler velocities, and $T\!F(\nu)$ is a transfer function of the atmosphere. 
The transfer function relates to the transmission of the wave amplitudes by the solar atmosphere as a function of frequency. It is proportional to the ratio of the velocity amplitude observable as a Doppler shift of the given line to the true amplitude. Its value is unity if the entire wave signal is observed as the Doppler shift of the spectral line throughout the atmosphere at a given frequency, whereas values smaller than unity represent some loss of the signal because of the extent in height of the spectral-line contribution functions. In general, the most affected are short-period (small-scale) fluctuations. A detailed time-dependent model of the atmosphere is needed to derive this value. We set this function equal to unity for all frequencies, which means that the observed Doppler signal of all waves at a given frequency is detected throughout the solar atmosphere \citepalias[cf.][]{sobotka2016}. This rough approximation may lead to underestimated acoustic-flux values at higher frequencies.

The group velocity for vertical energy transportation $\upsilon_{\text{gr}}$ is given by  
\begin{equation}
\begin{split}
\upsilon_{\rm gr} = c_{\rm s} \, {\sqrt{1 -(\nu_{\rm ac}/ \nu)^2}}, 
\label{equ:3}
\end{split}
\end{equation}
where $\nu_{\rm ac}$ = $\gamma g \cos \theta /(4 \pi c_{\rm s})$ is the acoustic cut-off frequency and $c_{\rm s}= \sqrt{\gamma P_{\rm g}/\rho}$ the sound speed.
Here, $\gamma$ is the adiabatic index equal to 5/3 for monoatomic gas, $g$ the surface gravity, and $\theta$ the magnetic-field inclination in the photosphere, which reduces the acoustic cut-off frequency \citep{Cally2006}. For $\nu_{\rm ac}$ we adopt the value $5.2 \cos \theta$~mHz. The total acoustic flux at all frequencies between $\nu_{\rm ac}$ and the maximum detectable frequency $\nu_{\rm max}$ is
\begin{equation}
\begin{split}
F_{\rm ac,tot} = \int_{\nu_{\rm ac}}^{\nu_{\rm max}} F_{\rm ac} \, {\rm d}\nu.
\label{equ:4}
\end{split}
\end{equation}
The values of gas pressure $P_{\text g}$ and density $\rho$ are taken in each pixel of the ROIs from the corresponding scaled model atmosphere at two reference heights $h_1$ and $h_2$. The deposited acoustic flux $\Delta F_{\rm ac}$ is the difference between the incoming acoustic energy flux at the lower reference height $h_1$ and the outgoing one at the upper reference height $h_2$,
\begin{equation}
\begin{split}
\Delta F_{\rm ac} = F_{\rm ac,tot}(h_1) -  F_{\rm ac,tot}(h_2).
\label{equ:5}
\end{split}
\end{equation}
A part of the incoming energy flux is dissipated in the chromosphere between $h_1$ and $h_2$ and likely converted into radiation, while the outgoing part continues to propagate higher in the atmosphere.


\section{Results}$\label{sec:Results}$

The best-matching semi-empirical models (see Section~\ref{sec:Modelatmosphere}) were assigned to all positions in ROIs of all data sets. Then the deposited acoustic fluxes were calculated for each position. The most important input parameters for the $\Delta F_{\rm ac}$ calculation are the Doppler velocity amplitudes, gas densities and pressures at the reference heights, and the magnetic-field inclination (Figure~\ref{fig:fig2}). Table~\ref{tab:table2} gives an overview of the first three quantities, where the velocity magnitudes are represented by the rms Doppler velocity and the gas densities and pressures by their mean values and standard deviations obtained by averaging over the corresponding ROIs.

\subsection{October 3 data set -- quiet Sun} \label{sec:Oct3}

The data set acquired on 2019 October 3 with the GST/FISS instrument includes a quiet-Sun region at the center of the solar disc. The lines Ca\,\textsc{ii} 854.2 nm and H$\alpha$ provide the information about two different ranges of heights, 1000--1400 km (middle chromosphere) and 1400--1800 km (upper chromosphere) -- see Section~\ref{sec:Modelatmosphere}. In the quiet-Sun atmosphere, the reference heights for the Ca\,\textsc{ii} wing, Ca\,\textsc{ii} center, and H$\alpha$ center are 1000, 1400, and 1800~km, respectively. The deposited acoustic fluxes $\Delta F_{\rm ac}$ and the total integrated radiative losses $L$ are computed for those height ranges.

\begin{figure}[t]\centering
\includegraphics[width=3.3in]{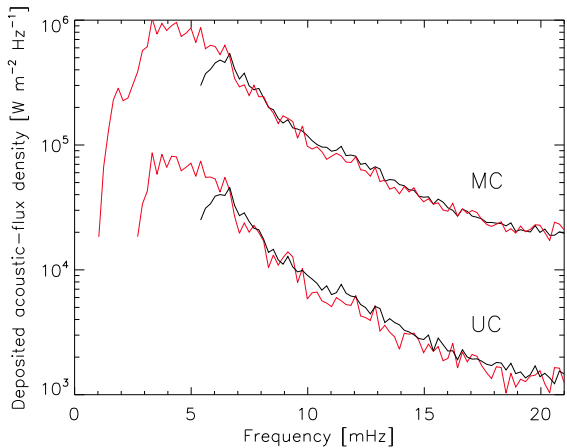}
\caption{Frequency distribution of the deposited acoustic-flux density averaged over magnetic ({\it red}) and non-magnetic ({\it black}) areas of the October 3 data set. MC and UC lines correspond to the height ranges 1000--1400 km (middle chromosphere) and 1400--1800 km (upper chromosphere), respectively.
\label{fig:fig5}}
\end{figure}

The frequency distribution of the deposited acoustic-flux density spatially averaged over the magnetic ($|B| > 125$~G, 11.6\,\% of ROI) and non-magnetic ($|B| < 125$~G, magnetic inclination set to zero, 88.4\,\% of ROI) areas is displayed in Figure~\ref{fig:fig5} for the two height ranges. The maximum contribution in non-magnetic areas is at 6--7~mHz and for frequencies higher than 14~mHz, the contribution is by an order of magnitude smaller than the maximum one. The presence of inclined magnetic fields adds a substantial contribution at frequencies below 5.2~mHz, increasing the total deposited acoustic fluxes in the middle and upper chromosphere by a factor of 2.3.

\begin{figure}[t]\centering
\includegraphics[width=3.3in]{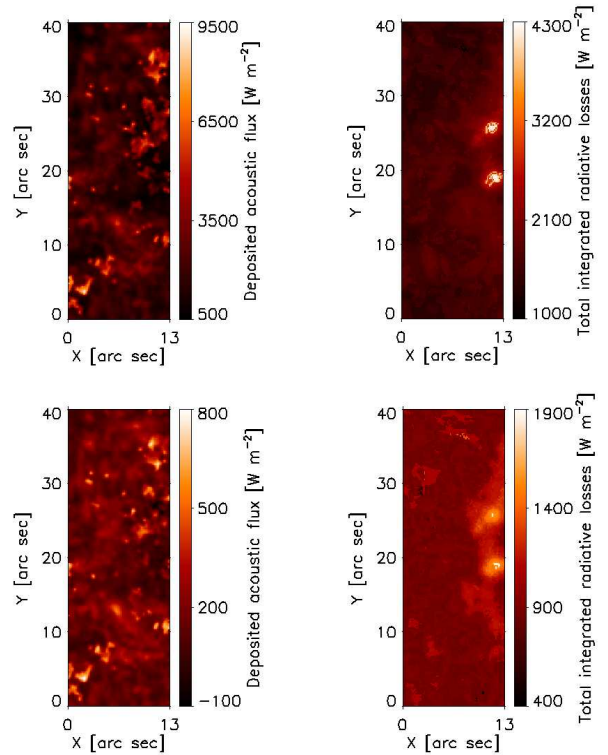}
\caption{The deposited acoustic energy flux ({\it left}) and the total integrated radiative losses ({\it right}) in the ROI of the October 3 data set. {\it Top}: $h = 1000$--1400 km (middle chromosphere), {\it bottom}: $h = 1400$--1800 km (upper chromosphere).
\label{fig:fig6}}
\end{figure}

\begin{figure}[t]\centering
\includegraphics[width=3.6in]{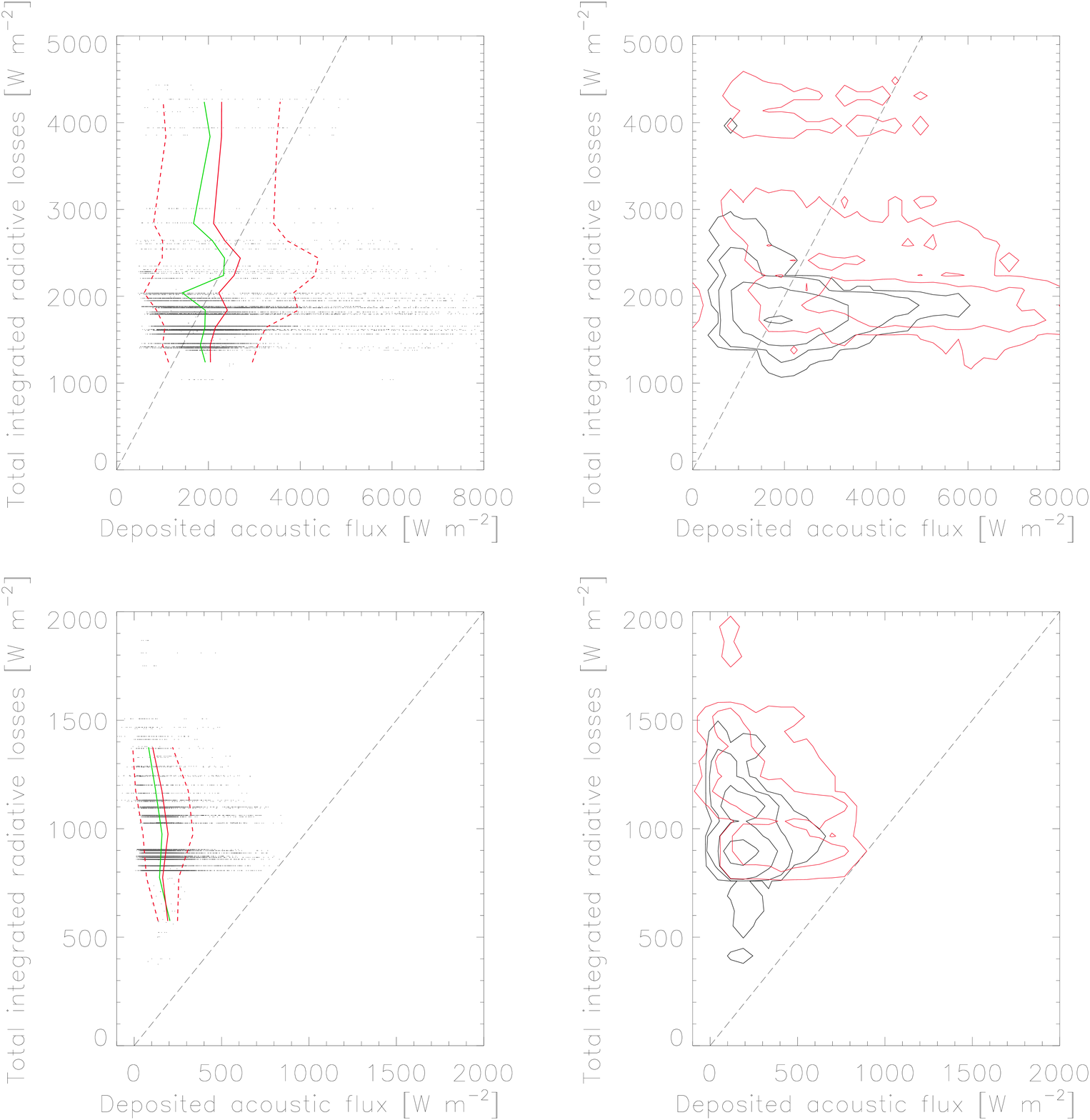}
\caption{{\it Left}: Scatter plots of total integrated radiative losses versus deposited acoustic flux in the ROI of the October 3 data set. {\it Top}: \mbox{$h = 1000$--1400 km}, {\it bottom}: $h = 1400$--1800 km. Solid lines show average ({\it red}) and median ({\it green}) values together with red dashed lines of $\pm 1 \sigma$. {\it Right}: Density contours of the scatter plots. Contours for magnetic ({\it red}) and non-magnetic ({\it black}) areas are plotted separately at density levels of 0.03, 0.3, 3, and 30\,\% of a common density maximum. Straight dashed lines represent the full balance of radiative losses by acoustic-flux deposit.
\label{fig:fig7}}
\end{figure}

The maps of the deposited acoustic flux and total integrated radiative losses are displayed in Figure~\ref{fig:fig6}. The scatter and density (2D histogram) plots of the $\Delta F_{\rm ac}$ to $L$ pixel-by-pixel comparison are shown in Figure~\ref{fig:fig7}.
The spatial distribution of $\Delta F_{\rm ac}$ is strongly intermittent, as already noted by \cite{Bello2010b}, and we observe a large scatter of $\Delta F_{\rm ac}$ versus $L$. For this reason, we have to analyze the contribution of the deposited acoustic flux to radiative losses statistically, using the mean, median, and standard deviation of $\Delta F_{\rm ac}$. These statistical values are calculated in 100~W\,m$^{-2}$ wide bins of the $L$ histogram and each bin must contain at least 50 points. In Figure~\ref{fig:fig7} {\it left}, the red and green solid lines represent the bin-averaged $\overline{\Delta F_{\rm ac}}$ and median values, respectively, and the red dashed lines set the boundaries of the $\pm 1 \sigma$ range that characterize the scatter of individual points in each bin. In the density plots (Figure~\ref{fig:fig7} {\it right}), pixels of magnetic and non-magnetic areas are separated and their densities are normalized to the common histogram maximum of the two areas. The bin size of the 2D histogram is 30~W\,m$^{-2} \times 30$~W\,m$^{-2}$. Contours are plotted at four density levels with a step of one order of magnitude, beginning at 0.03\,\% of the maximum.

In the middle chromosphere, the total radiative losses integrated over the 400 km thick layer are $1000 < L < 2600$~W~m$^{-2}$ in the quiet area and reach 4300~W~m$^{-2}$ in two bright points connected with magnetic elements (bottom panels in Figure~\ref{fig:fig2} and right panels in Figure~\ref{fig:fig6}). The ratio of the mean deposited acoustic flux to the radiative losses $\overline{\Delta F_{\rm ac}}/L$ is between 0.9 and 1.3 in the quiet area, so that the acoustic energy flux balances the energy released by radiation (top panels in Figure~\ref{fig:fig7}). However, the contribution of $\overline{\Delta F_{\rm ac}}$ to $L$ in the two bright points is only 60\,\%. 

In the upper chromosphere at the heights 1400--1800 km, the radiative losses ($400 < L < 1900$~W~m$^{-2}$) are larger than the deposited acoustic flux for all the points in the scatter plot (bottom panels in Figure \ref{fig:fig7}) and \mbox{$\overline{\Delta F_{\rm ac}}/L \simeq 0.2$.} This means that the deposited acoustic flux is insufficient to balance the radiative losses and maintain the semi-empirical temperature at these heights. Negative values of $\Delta F_{\rm ac}$ appearing at some scarce locations in the ROI are caused by the limited accuracy of our method (see Section~\ref{sec:Modelatmosphere}).

\begin{table*}\centering
\caption{October 3 data set: Spatial averages of incoming acoustic flux $F_{h1}$, its part $\Delta F_{\rm ac}$ deposited between the reference heights $h_1$ and $h_2$, and corresponding radiative losses $L$.}
\label{tab:table3}
\begin{tabular}{lcccccc}
\hline\hline
Quiet-Sun region & $h_1 - h_2$~[km] & $F_{h1}$~[W\,m$^{-2}$] &  $\Delta F_{\rm ac}$~[W\,m$^{-2}$] & $\Delta F_{\rm ac} /F_{h1}$ & $L$~[W\,m$^{-2}$] & $\Delta F_{\rm ac} /L$ \\
\hline
Magnetic area, & 1000--1400 & $4960 \pm 2010$ & $4440 \pm 1814$ & $0.90 \pm 0.01$ & $2087 \pm 596$ & $2.13 \pm 1.43$  \\
11.6\,\% of ROI & 1400--1800 & $520 \pm 202$ & $321 \pm 191$ & $0.62 \pm 0.20$ & $1057 \pm 203$ & $0.30 \pm 0.24$  \\
\hline
Non-magnetic area, & 1000--1400 & $2085 \pm 744$ & $1871 \pm 673$ & $0.90 \pm 0.01$ & $1689 \pm 180$ & $1.11 \pm 0.42$ \\
88.4\,\% of ROI & 1400--1800 & $214 \pm 74$ & $147 \pm 65$ & $0.69 \pm 0.10$ & $888 \pm 92$ & $0.17 \pm 0.08$ \\
\hline
\end{tabular}
\tablefoot{Standard deviations represent a scatter of individual values in the areas.}
\end{table*}

To estimate the fraction of the incoming acoustic energy flux $F_{h1} = F_{\rm ac,tot}(h_1)$ deposited in a layer between the reference heights $h_1$ and $h_2$, we use the spatially-averaged values of $F_{h1}$ and $\Delta F_{\rm ac}$ in the magnetic and non-magnetic areas in the ROI. The results are shown in Table~\ref{tab:table3} together with the corresponding spatially averaged values $L$ of the total integrated radiative losses and the ratio $\Delta F_{\rm ac} / L$.
Standard deviations characterize a large scatter of values corresponding to individual pixels in these areas. The incoming acoustic flux at 1000~km is by a factor of 2.4 larger in the magnetic area than in the non-magnetic one. In both types of areas, 90\,\% of the incoming flux is deposited between the heights 1000--1400~km, which is sufficient to balance the radiative losses. A fraction of 0.6--0.7 of the remaining 10\,\% of the acoustic flux passing through the height of 1400~km is deposited below 1800~km, but its contribution to the radiative losses is small. Acoustic fluxes of 200~W~m$^{-2}$ in the magnetic area and 70~W~m$^{-2}$ in the non-magnetic one pass to layers above 1800~km.

\subsection{December 11 and June 6 data sets -- weak active regions} \label{sec:Dec11}

Two weak active regions including plages and small pores were observed on 2018 December 11 and 2019 June 6 with the VTT echelle spectrograph in the hydrogen lines H$\alpha$ and H$\beta$ (see Section~\ref{sec:observation}). 
The Doppler velocities observed in the line center of H$\alpha$ (reference height 1900~km in active regions, see Section~\ref{sec:Modelatmosphere}) and at the distance $\Delta\lambda$ = $\pm$18 pm from the line center of H$\beta$ (reference height 1600~km) are used to calculate acoustic energy fluxes (Section~\ref{sec:Deposited}).

\begin{figure}[t]\centering
\includegraphics[width=3.3in]{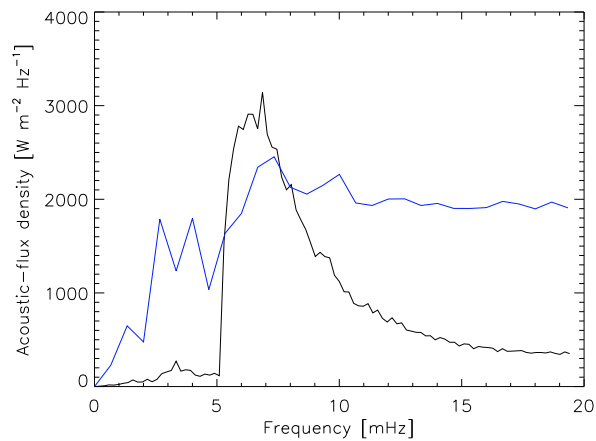}
\caption{Frequency distribution of the incoming acoustic-flux density at $h_1 =1600$~km averaged over the whole ROIs (magnetic plus non-magnetic areas) of December 11 ({\it blue}) and June 6 ({\it black}).
\label{fig:fig8}}
\end{figure}

The amplitudes of H$\beta$ velocities are quite small in comparison with the other chromospheric lines (see Table~\ref{tab:table2}). This results in very small mean incoming acoustic fluxes $F_{\rm ac,tot}(1600\, \rm km)$, which, however, differ for the December and June data sets, being equal to 40 and 15~W\,m$^{-2}$, respectively. This difference can be explained by different frequency distributions plotted in Figure~\ref{fig:fig8}. On June~6, the shape of the frequency distribution is similar to that on October 3 (Figure~\ref{fig:fig5}), while the December 11 data exhibit significant contributions at frequencies in the ranges \mbox{1--5 mHz} (thanks to the large fraction of magnetised area) and \mbox{8--20 mHz}, induced probably by seeing, which was not sufficiently good during the December 11 run.

\begin{figure}[t]\centering
\includegraphics[width=3.3in]{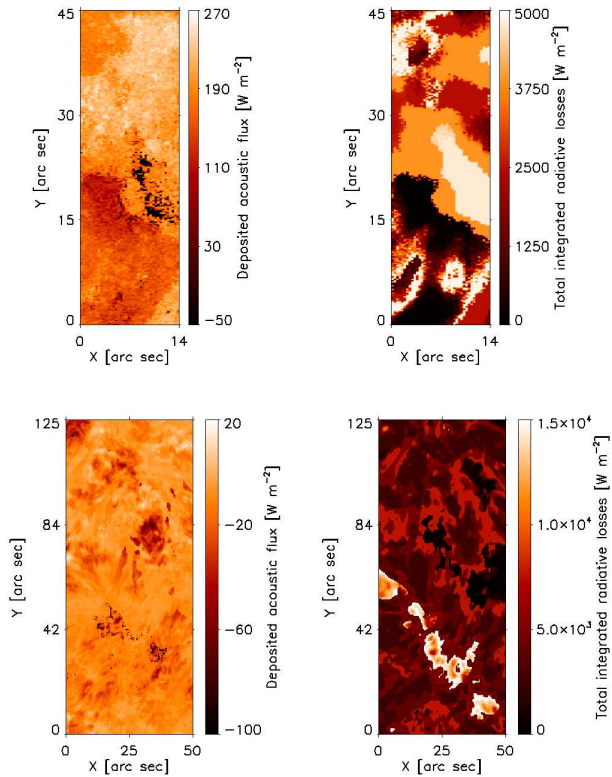}
\caption{The deposited acoustic energy flux ({\it left}) and the total integrated radiative losses ({\it right}) for the range \mbox{$h$=1600--1900 km} (upper chromosphere) in the ROIs of December 11 ({\it top}) and June 6 ({\it bottom}).
\label{fig:fig9}}
\end{figure}

The deposited acoustic fluxes $\Delta F_{\rm ac}$ and total integrated radiative losses $L$ are computed and integrated for the range \mbox{$h$ = 1600--1900 km} in the upper chromosphere. The maps of $\Delta F_{\rm ac}$ and $L$ for both ROIs are displayed in Figure~\ref{fig:fig9}. Enhanced radiative losses of the order $10^4$~W\,m$^{-2}$ are observed in the central part of the plage on June~6. This area is characterized by hot atmospheric models, obtained by scaling the initial model VAL~F (see Figure~\ref{fig:fig3}), for which a strong contribution of hydrogen Lyman-$\alpha$ to $L$ at $h > 1800$ km becomes important.

\begin{figure}\centering
\includegraphics[width=3.6in]{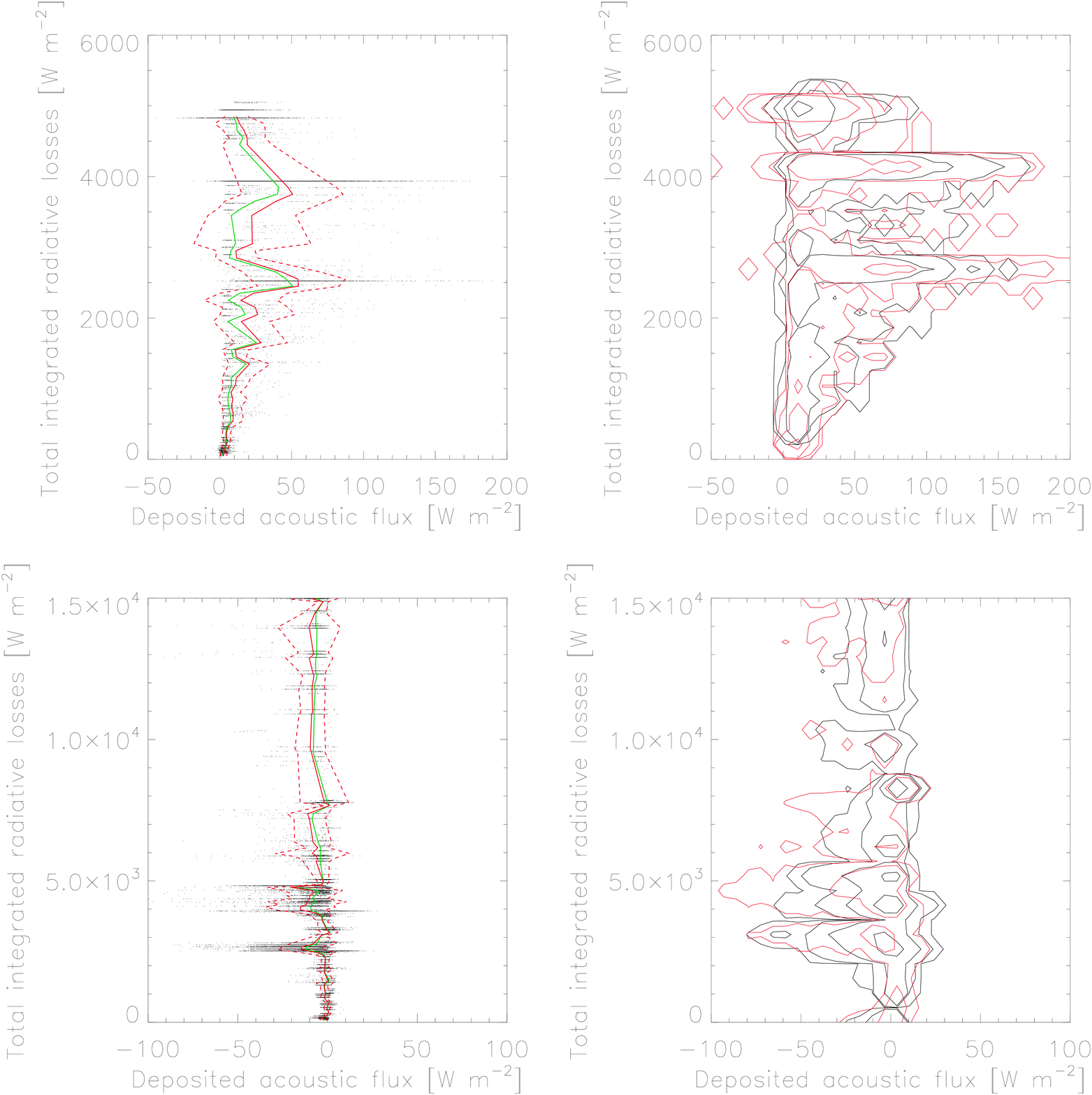}
\caption{{\it left}: Scatter plots of total integrated radiative losses versus deposited acoustic flux for the range \mbox{$h$=1600--1900 km} in the ROIs of December 11 ({\it top}) and June 6 ({\it bottom}). Solid lines show average ({\it red}) and median ({\it green}) values together with red dashed lines of $\pm 1 \sigma$. {\it Right}: Density contours of the scatter plots. Contours for magnetic ({\it red}) and non-magnetic ({\it black}) areas are plotted separately at density levels of 0.03, 0.3, 3, and 30\,\% of a common density maximum.
\label{fig:fig10}}
\end{figure}

The scatter and density plots comparing $\Delta F_{\rm ac}$ and $L$ for both ROIs are shown in Figure~\ref{fig:fig10}. The mean, median, and standard deviation of the deposited acoustic flux are calculated in the same way as in Section~\ref{sec:Oct3}. For all plotted points, the deposited acoustic flux is mostly in the range from $-50$\,W\,m$^{-2}$ to 150\,W\,m$^{-2}$ and it is very small compared to the radiative losses. Its mean values are approximately 30~W\,m$^{-2}$ for the December 11 data set and zero for the June 6 one.
The small positive mean value of the December flux is a consequence of the high-frequency contribution induced probably by the seeing. In the density plots, magnetic and non-magnetic areas are distinguished. The densities are calculated and the contour levels are set in the same way as for the October 3 data set (Section~\ref{sec:Oct3}). The magnetic areas of the December and June data sets include 33.0\,\% and 3.9\,\% of pixels in the ROIs, respectively.

A co-temporal measurement of the magnetic-field vector obtained with GRIS at the GREGOR telescope (Section~\ref{sec:Dataset}) was used alternatively to re-compute the deposited acoustic fluxes of the December 11 data set. Because the spatial resolution of the GRIS magnetic-inclination map was by a factor of two better than that of HMI, the scatter of $\Delta F_{\rm ac}$ values was reduced but the mean and median values did not change.

Thus, the contribution of the deposited acoustic flux to the radiative losses is practically zero in the layer between \mbox{$h$ = 1600--1900 km}. This confirms the finding of Section~\ref{sec:Oct3} that $\Delta F_{\rm ac}$ cannot balance $L$ in the upper chromosphere.
We expect that the scatter of deposited acoustic-flux values around zero in the December and June data sets is caused by a limited accuracy in the determination of incoming and outgoing acoustic fluxes. The fact that Doppler velocities are sometimes measured at heights different from the reference ones (Section~\ref{sec:Modelatmosphere}), introduces the relative error of 15\,\% in both quantities. Moreover, when the inner wings of H$\beta$ are affected by a photospheric contribution, the error of the incoming acoustic flux increases to 30\,\%.


\section{Discussion and conclusions \label{sec:conclusion}}

We studied the dissipation of (magneto)acoustic waves through different layers of the solar chromosphere using the VTT/echelle spectrograph and GST/FISS multi-line observations in H$\alpha$, H$\beta$ and Ca\,\textsc{ii} 854.2~nm. The deposited acoustic energy fluxes were quantitatively compared with the total radiative losses in quiet-Sun and weak active regions in the central zone of the solar disc.

We utilized time series of Doppler velocities measured in the line centers and wings (see Section~\ref{sec:Deposited}) to derive the deposited acoustic fluxes. The maximum detectable frequency of waves was 20 mHz. The calculation of radiative losses was based on a set of non-LTE 1D hydrostatic semi-empirical models assigned to different positions in the ROIs according to the best match of synthetic profiles to time-averaged observed profiles. The models were obtained by scaling the temperatures and column masses of the initial models VAL B--F.

We have demonstrated that the radiative losses can be fully balanced by the deposited acoustic energy flux in the middle chromosphere \mbox{($h$ = 1000--1400 km)} of a quiet-Sun region observed in the October 3 data set. Seemingly, this contradicts the result of \citetalias{Abbasvand2020}, stating that the deposited acoustic flux contributes to the radiative losses only by 30--50\,\% in quiet areas. The discrepancy can be explained by the fact that the area considered as quiet Sun in \citetalias{Abbasvand2020} was close to a plage and fell within its extended canopy region. Thus, the effect of magnetic shadows \citep{Vecchio2007,Konto2010}, related to the elevated magnetic field of the canopy, reduced the oscillatory power and the deposited acoustic flux. The quiet-Sun region studied in the current work was far from any plage or pore, so that the propagation of waves in the chromosphere was less affected by canopy fields.

Radiative losses $L > 4000$~W\,m$^{-2}$ in small bright chromospheric features related to magnetic elements in the quiet region were larger than the deposited acoustic flux by a factor of 1.7 on average. The enhancement of deposited acoustic flux at frequencies below 5.2~mHz in moderately inclined magnetic field of these features did not compensate the energy released by radiation.

We have also shown that in the upper chromosphere, the incoming and deposited acoustic fluxes are very small compared to the radiative losses. In quiet-Sun region, the acoustic-flux deposit contributes only by about 20\,\% to the radiative losses between the heights 1400~km and 1800~km and in the weak active regions (data sets of December 11 and June 6), its contribution is practically equal to zero in the range \mbox{1600--1900 km}.

This finding may be explained by the fact that the acoustic energy flux at a given height is proportional to the density of gas $\rho$, power density of Doppler velocities $P_\upsilon$, and the group velocity $\upsilon_{\rm gr}$ (Equation~\ref{equ:2}). Of these three quantities, the change of $\rho$ with height is the most important one. Densities in the middle chromosphere are by an order of magnitude higher than those in the upper chromosphere, while $P_\upsilon \sim \sigma_{\upsilon}^2$ is approximately of the same order (see Table~\ref{tab:table2}) and $\upsilon_{\rm gr}$ (Equation~\ref{equ:3}) may change by a factor of 3 at maximum. Thus, it is mainly the density that determines the amounts of the incoming and deposited acoustic fluxes, which are considerable in the middle chromosphere \mbox{($h$ = 1000--1400 km)} but negligible in the upper layers.

We can conclude that the acoustic energy flux is efficiently deposited in the middle chromosphere, where the density of gas is sufficiently high. The deposited acoustic flux is a major contributor to the radiative losses of this layer in quiet-Sun regions and in plages \citepalias[see][]{Abbasvand2020}. However, the energy transported by (magneto)acoustic waves can be reduced by magnetic shadows of canopy regions around plages and pores. In the upper chromosphere, the incoming -- and thus deposited -- acoustic flux cannot balance the radiative losses, which can be large in active regions thanks to the radiation in the hydrogen Lyman-$\alpha$ line formed at the top of the chromosphere. This means that other heating mechanisms have to act in the upper chromosphere.

We would like to emphasize that our results are not inconsistent with the previous studies, where many (e.g., \citeauthor{Fossumetal2005} \citeyear{Fossumetal2005}; \citeauthor{Beck2009} \citeyear{Beck2009}, or see a thorough discussion in Section 4.1 of \citeauthor{Jess2015} \citeyear{Jess2015}) showed that the (magneto)acoustic waves are not sufficient to explain the heating of the chromosphere.
In the present work we only studied two separate layers of the chromosphere available to us in our observations. We have no information about the relations between the deposited acoustic flux and radiative losses in the lower chromosphere. Thus, we do not attempt to draw any conclusions regarding the energy budget of the chromosphere as a whole. We point out that for some layers the heating by (magneto)acoustic waves may be sufficient, whereas in other layers it is insufficient. This fact itself draws a need for an additional source of energy. It is interesting to note that a similar conclusion was drawn by \cite{Fawzy2002} in the case of chromospheres of six late-type main-sequence stars, for which high-quality spectroscopic observations were available.

We are aware that our stationary approach (time-averaged acoustic fluxes and 1D semi-empirical models) is only a rough approximation to the real time-dependent physical conditions in the chromosphere. For this reason, we consider as meaningful just the statistical comparison of the deposited acoustic fluxes to the radiative losses, trying to provide a general estimate of the role of (magneto)acoustic waves in the heating of long-lived chromospheric structures.


\begin{acknowledgements}
We thank the anonymous referee for valuable comments and suggestions.
This work was supported by the Czech Science Foundation and Deutsche Forschungsgemeinschaft under the common grant 18-08097J---DE 787/5-1 and the institutional support ASU:67985815 of the Czech Academy of Sciences. 
The observations were partly supported by the project SOLARNET that has received funding from the European Union Horizon 2020 research and innovation programme under grant agreement no. 824135 and by the Czech Science Foundation grant 18-06319S.
J.K. acknowledges the project VEGA 2/0048/20, M.G.R. the Charles University project GA UK No. 204119, and D.K the project GA UK No. 532217.
BBSO operation is supported by the New Jersey Institute of Technology and US NSF AGS-1821294 grant. GST operation is partly supported by the Korea Astronomy and Space Science Institute, the Seoul National University, and the Key Laboratory of Solar Activities of Chinese Academy of Sciences (CAS) and the Operation, Maintenance and Upgrading Fund of CAS for Astronomical Telescopes and Facility Instruments.
We are grateful to N. Gorceix, C. Plymate, J. Varsik, and the BBSO staff for operating the instrument, observing the data, and performing the basic data reduction.
The 1.5-m GREGOR solar telescope was built by a German consortium under the leadership of the Leibniz-Institut f\"ur Sonnenphysik in Freiburg (KIS) with the Leibniz-Institut f\"ur Astrophysik Potsdam (AIP), the Institut f\"ur Astrophysik G\"ottingen, the Max-Planck-Institut f\"ur Sonnensystemforschung in G\"ottingen (MPS) as partners, and with contributions by the Instituto de Astrof{\'\i}sica de Canarias (IAC) and the Astronomical Institute of the Czech Academy of Sciences.
The Vacuum Tower Telescope at the Spanish Observatorio del Teide of IAC is operated by the German consortium of KIS, AIP, and MPS.
\end{acknowledgements}


\bibliographystyle{aa}
\bibliography{bibliography1}

\end{document}